\documentstyle[aps,eqsecnum,epsfig]{revtex}
\tighten
\begin{document}
\draft

\twocolumn[\columnwidth\textwidth\csname@twocolumnfalse\endcsname

\title{Nuclear Skins and Halos in the Mean-Field Theory}

\author  {S. Mizutori,$^{1,2,3}$
J. Dobaczewski,$^{1,2,4}$
G.A. Lalazissis,$^{1,2,5}$
W. Nazarewicz,$^{1,2,6}$ and
P.-G. Reinhard$^{1,7}$
}

\address {$^1$Joint Institute for Heavy Ion Research,
              Oak Ridge National Laboratory,
              P.O. Box 2008, Oak Ridge,   Tennessee 37831}
\address {$^2$Department of Physics and Astronomy, University
              of Tennessee Knoxville,  Tennessee 37996}
\address {$^3$Department of Physics, Osaka University,
        Toyonaka, Osaka 560-0043, Japan}
\address {$^4$Institute of Theoretical Physics, Warsaw University,
              ul. Ho\.za 69, PL-00681, Warsaw, Poland}
\address {$^5$Physics Department, Aristotle University of
 Thessaloniki, 54006 Greece}
\address {$^6$Physics Division, Oak Ridge National Laboratory,
              P.O. Box 2008, Oak Ridge,   Tennessee 37831}
\address {$^7$Institut f\"ur Theoretische Physik, Universit\"at Erlangen
              Staudtstr.\ 7, D-91058 Erlangen, Germany}

\maketitle
\begin{abstract}
Nuclei with large neutron-to-proton ratios
 have neutron skins, which
manifest themselves in an excess of neutrons at distances
greater than the radius of the proton distribution. In addition, some drip-line nuclei
develop  very extended
halo structures. The neutron halo is a threshold effect;
it  appears when the valence neutrons
occupy weakly bound orbits.
In this study, nuclear skins and halos are analyzed within the
self-consistent Skyrme-Hartree-Fock-Bogoliubov and
relativistic Hartree-Bogoliubov theories for spherical shapes.
It is demonstrated that
skins, halos, and surface thickness can be analyzed  in a
model-independent way in terms of  nucleonic density form factors.
Such an analysis allows for defining a quantitative measure of the
halo size. The
systematic behavior of skins, halos, and surface thickness
in even-even nuclei is discussed.
\end{abstract}


\addvspace{5mm}]

\narrowtext

\section{Introduction}\label{intro}

One of the main  frontiers of nuclear science today is
the physics of radioactive nuclear beams (RNB).
Experiments with beams of unstable nuclei make it possible
 to look
closely into many unexplored regions of the periodic chart
and many unexplored
 aspects of the nuclear many-body problem
\cite{[Naz96a],[ISOL],%
[Dob97a],[Mue99],[Tan99]}.

Prospects for new physics, especially on the neutron-rich side
of the beta-stability valley, have generated  considerable
excitement in the low-energy nuclear physics community.
Neutron-rich nuclei
offer an opportunity to study the wealth of phenomena
associated with the closeness of the particle threshold:
particle emission (ionization to the continuum) and
characteristic behavior of cross sections \cite{[Wig48],[Fan61]},
 existence of soft
collective modes  and low-lying transition strength
\cite{[Uch85],[Fay91],[Yok95],[Sag95],[Ham96],[Lan99]},
and dramatic changes in shell structure and
various nuclear properties in the sub-threshold regime
\cite{[Ton78],[Dob94],[Dob96]}.

A very interesting aspect of
nuclei far from stability is an increase in their radial
dimension with decreasing particle separation energy
\cite{[Rii92],[Mue93],[Rii94],[Han95],[Tan96]}. Extreme cases
are halo nuclei -- loosely bound few-body systems with about
thrice more neutrons than protons.  The halo region is a zone of
weak binding in which quantum effects play a critical role in
distributing nuclear density in regions not classically allowed.

Halo nuclei, with their intricate topologies,
 are symbols of RNB physics. The very weak
binding of the outermost neutrons leading to a
 rather good decoupling of halo from the core simplifies
many  aspects of underlying nuclear structure and reaction mechanism.
Theoretically, the weak binding and corresponding
closeness of the particle continuum, together with
the need for the explicit treatment of few-body dynamics,
makes the  subject of halos both extremely interesting
and difficult \cite{[Dob97a]}.

In the heavy
neutron-rich  nuclei, where  the concept of mean field
is better applicable, an {\it a priori}
separation into core and halo
nucleons  seems less justified.
However,
the fact that there are far more neutrons than protons
in these nuclei implies
the existence of the neutron skin
(i.e., an excess of neutrons at large distances). In addition,
in neutron-rich weakly bound nuclei, one expects to see
both the skin and the halo.

There is no consensus in the literature on  how to
define and parametrize skins and halos.
A quantity  which is often employed to characterize
the spatial extension of neutron density is the
difference between neutron and proton root mean square (rms) radii:
\begin{equation}\label{rnp1}
\Delta R_{np} \equiv \langle r_n^2\rangle - \langle r_p^2\rangle.
\end{equation}
In normal nuclei, this quantity is known to vary between 0.1--0.2\,fm
\cite{[Hod71],[Bat89],[Kra91],[Dob96a]}, but it increases significantly
in neutron-rich systems due to {\em both} skin and halo effects.

Stimulated by recent experimental developments
mainly in light nuclei, where some
information on  $\Delta R_{np}$ has been obtained
\cite{[Suz95],[Chu96],[Suz98],[Boc98],[Mar99]}
(see also the recent studies based on the giant dipole \cite{[Kra94]}
and spin-dipole
\cite{[Kra99]} resonance data  and on  antiprotonic levels
\cite{[Lub98],[Sch99]}), many theoretical papers
with  a focus on radii of
neutron and proton density distributions
appeared
\cite{[Ham95],[Taj96],[Hor97],[Ots97],[Kit97],[Ren97],[Bar97],[Ren98],%
[Hof98],[Men98],[Men98a],[Len98],%
[War98],[Sto98],[Lal98],[Pat99],[Ren99],[Kny99],[Lal99],[Typ99],[Pol99]}.
(For earlier works, see papers quoted in Ref.~\cite{[Dob96a]}.)
In most cases,
theoretical studies were concerned with rms radii, and the skin
was usually discussed in terms of quantity (\ref{rnp1}).

Unfortunately,
the second moment of nucleonic density  (the rms radius) provides a very
limited characterization of the nucleonic distribution.  In particular,
since the parameter $\Delta R_{np}$
can be strongly  influenced by weakly bound
valence nucleons, i.e., by the shell structure,
 it is not able to properly describe the bulk radial
behavior of drip-line nuclei. A powerful tool that allows a more detailed description
is the Helm model, introduced in the context
of electron scattering experiments  \cite{[Hel56],[Ros67],[Rap70]}.
In this model, the diffraction radius and surface thickness extracted
from the density form factor are mainly sensitive
to the nucleonic distribution  in the surface region, and
they are practically independent of shell fluctuations in the nuclear interior
\cite{[Dur81],[Fri82],[Fri86],[Fri86a],[Spr92]}. The robustness
of the Helm model parameters, and their simple geometric interpretation, make
this model a very attractive tool when characterizing
  density distributions.

The main goal of this study is to apply the Helm model
to nucleonic densities calculated in the self-consistent mean-field theory.
In the first part of this paper,
it is demonstrated that by analyzing the nucleonic form factor one is able
to define, in a model-independent way, contributions to proton and neutron
radii coming from skins and halos. In the second part,  we perform
systematic calculations of skins and halos in spherical even-even nuclei and
discuss their dependence on the model employed.

The material contained in this study is organized as follows.
The analysis of nucleonic density based on
the Helm model is outlined in Sec.~\ref{helm}.
Section~\ref{models} discusses the
details of  Hartree-Fock-Bogoliubov (HFB) and 
relativistic Hartree-Bogoliubov (RHB) models employed.
The results of self-consistent calculations
for diffraction radii, surface thickness, skins, and halos
in spherical even-even nuclei
are discussed  in Sec.~\ref{results}, together
with the simple analysis based on the square-well model.
 Finally, Sec.~\ref{conclusions} contains
the main conclusions of this work.

\section{Spherical Helm Model}\label{helm}

A key feature of the nucleonic density is the rms radius
\begin{equation}\label{eq:rms}
  R_{\rm rms}
  \equiv
  \sqrt{\langle r^2\rangle}
  =
  \sqrt{\frac{\int d^3\bbox{r}\,r^2 \rho(\bbox{r})}
{\int d^3\bbox{r}\,\rho(\bbox{r})}}.
\end{equation}
Further characteristics  are best deduced from the corresponding
form factor
\begin{equation}\label{ff}
  F(\bbox{q})
  \equiv
  \int e^{i\bbox{qr}}\rho(\bbox{r})d^3r.
\end{equation}
For the spherical density distribution $\rho(r)$, the form factor
$F(q)$ is spherical and can be expressed in  the standard way:
\begin{equation}\label{ff1}
  F(q)
  =
  \int j_0(qr)\rho(r)r^2dr.
\end{equation}

There are various ways to characterize the basic pattern of the
nucleonic density. A choice that is straightforward and easy to use is
provided by
the Helm model \cite{[Hel56],[Ros67],[Rap70]}. Here,
nucleonic density is approximated by a convolution of
a sharp-surface density with radius $R_0$ with the Gaussian
profile, i.e,
\begin{equation}\label{sharp}
  \rho^{\rm(H)}(\bbox{r})
  =
  \int d^3\bbox{r}'\,f_{\rm G}(\bbox{r}-\bbox{r}')\rho_0 \Theta(R_0-|\bbox{r}'|),
\end{equation}
where
\begin{equation}\label{folding}
  f_{\rm G}(r)
  =
  \frac{1}{(2\pi)^{3/2}\sigma^3}e^{-\frac{r^2}{2\sigma^2}}.
\end{equation}
The radius in $R_0$ in Eq.~(\ref{sharp}) is the diffraction (box
equivalent) radius, and the folding width $\sigma$ in
Eq.~(\ref{folding}) models  the surface thickness. The density
$\rho_0$ is given by
\begin{equation}
\rho_0 = {3N \over {4\pi R_0^3}},
\end{equation}
hence the Helm density $\rho^{\rm(H)}$ is normalized to
the particle number $N$. The advantage of the Helm model
is that folding becomes a simple product in Fourier space,
thus yielding
\begin{equation}\label{F0}
  F^{\rm(H)}(q)
  =
  \frac{3}{R_0 q} j_1(qR_0) e^{-\frac{\sigma^2 q^2}{2}}.
\end{equation}
It is obvious that the first zero of $F^{\rm(H)}(q)$ is
uniquely related to the radius parameter $R_0$. The fit of
this model parameter is thus trivial. We simply relate $R_0$
to the first zero of the realistic form factor $F(q)$, i.e.,
\begin{equation}\label{R00}
  R_0
  =
  4.49341/q_1,
\end{equation}
where $q_1$ is the first zero of $F(q)$.
This means that  $R_0$ can be deduced from the diffraction minimum and this
is why it is called a diffraction radius
(or  the box-equivalent radius).
The
surface thickness parameter, $\sigma$, can be computed by
comparing  the values of microscopic and Helm form factors,
$F(q_m)$ and $F^{\rm(H)}(q_m)$,
at the first maximum $q_m$ of $F(q)$, which gives
\begin{equation}\label{thickness}
  \sigma^2=
  \frac{2}{q_m^2}\ln\frac{3R_0^2j_1(q_mR_0)}{R_0q_mF(q_m)}.
\end{equation}

We have now at our disposal three key parameters that characterize
 the microscopic nucleonic density:
the rms radius $R_{\rm rms}$ as defined in Eq.~(\ref{eq:rms}),
the diffraction radius $R_0$  from Eq.~(\ref{R00}), and
the surface thickness $\sigma$ given by Eq.~(\ref{thickness}).
The Helm model has only two independent parameters,  and
thus its rms radius can be expressed in terms of $R_0$ and $\sigma$:
\begin{equation}\label{msradius}
  R^{\rm(H)}_{\rm rms}
  =
  \sqrt{\frac{3}{5}\left(R_0^2 +5\sigma^2\right)}.
\end{equation}
Furthermore, it is more natural to discuss radii which pertain
to a geometrical size of the nucleus and,  therefore,
the prefactor $\sqrt{3/5}$ in Eq.~(\ref{msradius})
is rather inconvenient. We prefer to work with radii which we call
here the {\em geometric} radii, defined as
\begin{mathletters}\label{rmsradiusH}\begin{eqnarray}
  R_{\rm geom}
  &=&
  \sqrt{\frac{5}{3}}R_{\rm rms}, \label{rmsradiusHa} \\
  R_{\rm Helm}
  &=&
  \sqrt{\frac{5}{3}}R^{\rm(H)}_{\rm rms}
  =
  \sqrt{\left(R_0^2 +5\sigma^2\right)}. \label{rmsradiusHb}
\end{eqnarray}\end{mathletters}%
With this definition, the geometric radius becomes the box-equivalent radius
in the limit of a small surface thickness.

\section{Mean-field models}\label{models}

This section contains a very brief description of self-consistent
models applied in this work. Since these models are standard,
our discussion is limited to basic definitions and references.

\subsection{Hartree-Fock-Bogoliubov model}

The HFB approach is a variational method which uses
nonrelativistic
independent-quasiparticle states as trial wave functions \cite{[RS80]}.
An independent-quasiparticle state is defined as a vacuum of
quasiparticle operators which are linear combinations of
particle creation and annihilation operators. In this work,
instead of using the matrix representation corresponding to a
set of single-particle creation operators numbered by
the discrete index, we  use the spatial coordinate
representation \cite{[Bul80],[Dob84]}.  This is particularly useful when discussing
spatial properties of the variational wave functions and the
coupling to the particle continuum \cite{[Dob84],[Dob96]}.

In our HFB calculations, we employ the zero-range Skyrme interaction in
the particle-hole channel.
The total binding energy of a nucleus is obtained self-consistently
from the  energy functional \cite{[Que78]}:
\begin{eqnarray}
   {\cal E}  =  {\cal E}_{\rm kin}
             &+&{\cal E}_{Sk}
               +{\cal E}_{Sk,ls}\nonumber \\
             &+&{\cal E}_{C}
                +{\cal E}_{\rm pair}
               -{\cal E}_{\text{CM}}, \label{eq:Etot}
\end{eqnarray}
where
$ {\cal E}_{\rm kin}$ is the kinetic energy functional,
$   {\cal E}_{Sk} $ is the Skyrme functional,
$ {\cal E}_{Sk,ls}$ is the spin-orbit functional,
$ {\cal E}_C$ is the Coulomb energy (including the exchange term),
${\cal E}_{\rm pair}$ is the pairing energy, and
${\cal E}_{\text{CM}}$ is the center-of-mass correction.

In this work,  two Skyrme parametrizations are used:
SkP \cite{[Dob84]} and  SLy4
\cite{[Cha97]}. Both of
these selected forces  perform  well concerning
the total energies and
radii.  In particular, both SkP and SLy4 parametrizations have been
shown to
reproduce  long
isotopic sequences \cite{[Cha97],[Dob95c]}.

In the particle-particle channel, we use the SkP parametrization in the HFB/SkP
variant and the density-independent
volume delta interaction in the HFB/SLy4 variant. The
strength of the delta force was adjusted according to the prescription
given in Refs.~\cite{[Cha97],[Dob95c]}. For the details
of the calculations,  we refer the reader
to Refs.~\cite{[Dob84],[Dob96]}.

\subsection{Relativistic mean-field model}

Relativistic mean-field theory  has been proved to be a
powerful tool in describing various aspects of nuclear structure
\cite{[Rin96]}.
The model explicitly includes mesonic degrees of
freedom and describes the nucleons as Dirac particles.
Nucleons interact in a relativistic covariant manner
through the exchange of virtual mesons: the isoscalar
scalar $\sigma$-meson, the isoscalar vector $\omega$-meson,
and the isovector vector $\rho$-meson.  The model is based
on the one-boson exchange description of the
nucleon-nucleon interaction. The  starting point is the effective
Lagrangian density \cite{[Ser86],[Ser92]}
\begin{eqnarray}
{\cal L}&=&\bar\psi\left(i\gamma\cdot\partial-m\right)\psi
\nonumber\\
&&+\frac{1}{2}(\partial\sigma)^2-U(\sigma )
-\frac{1}{4}\Omega_{\mu\nu}\Omega^{\mu\nu}
+\frac{1}{2}m^2_\omega\omega^2\nonumber  \\
&&-\frac{1}{4}{\vec{\rm R}}_{\mu\nu}{\vec{\rm R}}^{\mu\nu}
+\frac{1}{2}m^2_\rho\vec\rho^{\,2}
-\frac{1}{4}{\rm F}_{\mu\nu}{\rm F}^{\mu\nu}
-g_\sigma\bar\psi\sigma\psi
\nonumber\\
&&-g_\omega\bar\psi\gamma\cdot\omega\psi-
g_\rho\bar\psi\gamma\cdot\vec\rho\vec\tau\psi -
e\bar\psi\gamma\cdot A {\textstyle{\frac{(1-\tau_3)}{2}}}\psi,
\label{lagrangian}
\end{eqnarray}
where
\begin{equation}
  U(\sigma)
  =
  \frac{1}{2}m^2_\sigma\sigma^2+\frac{1}{3}g_2\sigma^3+
  \frac{1}{4}g_3\sigma^4.
\end{equation}
Vectors in isospin space are denoted by arrows.
(Vectors in three-dimensional coordinate space are always
indicated by bold-faced symbols.)  The Dirac
spinor $\psi$ represents the nucleon with mass $m$, and $m_\sigma$,
$m_\omega$, and $m_\rho$ are the masses of the $\sigma$-meson, the
$\omega$-meson, and the $\rho$-meson, respectively. The meson-nucleon
coupling constants, $g_\sigma$, $g_\omega$, and $g_\rho$, and unknown
meson masses are parameters adjusted to fit nuclear matter data and
some static properties of finite nuclei. $U(\sigma)$ denotes the
non-linear $\sigma$ self-interaction \cite{[Bog77]} and
$\Omega^{\mu\nu}$, $\vec R^{\mu\nu}$, and $F^{\mu\nu}$ are field
tensors.

For the purpose of  the present
study, we choose  two RMF
parameterizations:   NL3 \cite{[Lal97]} and 
NL-SH \cite{[Sha93]}. The force NL3 stems from a fit including exotic 
nuclei, neutron radii, and information on giant resonances.
The  NL-SH parametrization
was fitted with a bias toward isotopic trends and it
 also uses information on neutron radii.

The relativistic extension of the HFB theory was introduced
in Ref.~\cite{[Kuc91]}. In the Hartree approximation for
the self-consistent mean field, one obtains  the RHB
 equations which are  solved self-consistently
in coordinate space by discretization on
the finite element mesh \cite{[Poe97]}.
 The spatial components,
$\mbox{\boldmath $\omega$, $\rho_3$}$, and ${\bf  A}$
vanish due to the time-reversal symmetry. Because of
charge conservation, only the third component of the
isovector rho meson contributes.
In the present investigation, the pairing interaction
has been approximated by a phenomenological finite-range Gogny force
with the  D1S parameter set \cite{[Ber84a]}.
This force has been  adjusted to the pairing
properties of finite nuclei all over the periodic table.

\section{Results and discussion}\label{results}

\subsection{Skins and halos in spherical heavy nuclei}

The Helm-model  characteristics of
calculated  density distributions are obtained from
the microscopic form factors (\ref{ff1}).
Figure~\ref{rhoF(Sn)} shows the
neutron densities for $^{120,150,170}$Sn
calculated in the RHB/NL3
 model and the
corresponding form factors.
The positions of the  first and second  zero
of the form factor (indicated by arrows) decrease
gradually with  neutron number
 reflecting  the steady increase of the
neutron radius [see Eq.~(\ref{R00})].
The zeros  of $F(q)$ are regularly spaced and the ratio of
$q_2/q_1$ is very close to
the ratio of the first two zeros of the spherical Bessel
function $j_1$.
It is also seen that, in the considered range of $q$,
 the envelope
of $q|F(q)|$ is practically constant \cite{[Dre74]}. All of these
observations confirm  that in the region of
low-$q$ values shown in Fig.~\ref{rhoF(Sn)},
the ``model-independent" analysis of  theoretical
density distributions, according to Ref.~\cite{[Fri82]},
can safely be performed.

Our analysis of neutron and proton radii in the Sn
isotopes is summarized in Fig.~\ref{RadSnall}. The most interesting
observation is that for the isotopes with $N$$\leq$82, the neutron geometric
radius $R_{\rm geom}$ (\ref{rmsradiusHa}) is very close to the
Helm radius $R_{\rm Helm}$ (\ref{rmsradiusHb}).
On the other hand, for nuclei heavier than $^{132}$Sn, the
former is appreciably greater than the latter.
This behavior suggests that the difference between $R_{\rm geom}$
and $R_{\rm Helm}$ is related to the size of the neutron
separation energy. Indeed, for $N$$\le$132, the two-neutron separation
energy is $S_{2n}$$>$12\,MeV, and it drops to a few MeV around
$N$=100. Due to the weaker binding, the neutron distributions in the
very heavy tin isotopes have larger spatial extensions, and this
increases $R_{\rm geom}$ dramatically due to the weight $r^2$ in
Eq.~(\ref{eq:rms}). On the other hand, the form factor at
intermediate values of $q$ is almost independent of
the asymptotic tail of the density
distribution. Therefore, the radius parameters deduced
from the form factor,
$R_0$ and $R_{\rm Helm}$, show a less dramatic growth.

Guided by this observation, we introduce the halo parameter as the
difference
\begin{equation}\label{rhalo}
  \delta{R}_{\rm halo}
  \equiv
  R_{\rm geom} -  R_{\rm Helm}.
\end{equation}
Such a halo parameter is indicated in Fig.~\ref{RadSnall}, where it
shows the size of the neutron halo in neutron-rich tin isotopes.
It should be noted that a halo may also be defined through the higher
radial moments, e.g., $\langle r^4 \rangle$. We have checked,
however, that
other definitions do not have any advantage over the simplest
prescription (\ref{rhalo}).

In
contrast to the neutron halo parameter $\delta{R}_{\rm halo}$(n), for
protons the
value of $\delta{R}_{\rm halo}$(p) turns out to stay very
close to zero, i.e.,
\begin{equation}\label{rhalop}
  R_{\rm geom}{\rm(p)}
  \approx
  R_{\rm Helm}{\rm(p)},
\end{equation}
such that one cannot  easily resolve the difference of
$R_{\rm geom}$(p) and $R_{\rm Helm}$(p) in the plot. This
reflects the fact that protons are always very well localized in the
nuclear interior by the Coulomb barrier, and they are very well bound.
(The two-proton separation energy increases from
$S_{2p}$$\approx$4\,MeV in $^{100}$Sn to $\approx$28\,MeV in
$^{130}$Sn.)

Figure~\ref{rhor4} shows the calculated neutron densities
for the tin isotopes multiplied by $r^4$. (The area under $\rho r^4$ is
proportional to  $\langle r^2\rangle$.)
It is seen that the large value
of ${R}_{\rm geom}$ in HFB/SLy4 can be attributed to the presence
of a hump in $\rho r^4$ at $r$$\approx$15\ fm. In the
particular representation of Fig.~\ref{rhor4}, it is possible, at least
in the  HFB/SLy4 model, to see rather clearly the decoupling of nuclear density into
the  ``core" and ``halo"  parts. This decoupling is much
weaker in the  HFB/SkP model, and it is almost invisible in
RHB/NL3.

While
the halo is a property of neutrons or protons,
the neutron skin depends on the difference between neutron and proton radii,
and
thus it is more difficult to quantify.
 Indeed, since
several definitions of a radius have been employed in this work,
one can introduce various
parameters  reflecting the
neutron-proton radius difference, e.g.,
\begin{mathletters}\label{rskin}\begin{eqnarray}
  \Delta R_0
&  = &
  R_0{\rm(n)}-R_0{\rm(p)},\label{rskina}\\
  \Delta R_{\rm Helm}
&  = &
  R_{\rm Helm}{\rm(n)}-R_{\rm Helm}{\rm(p)},\label{rskinb}\\
  \Delta R_{\rm geom}
 & = &
  R_{\rm geom}{\rm(n)}-R_{\rm geom}{\rm(p)}.\label{rskinc}
\end{eqnarray}\end{mathletters}%
These three definitions are displayed in Fig.~\ref{skin_vdefs}
for the Sn isotopes calculated in the HFB/SLy4 model.

According to the discussion above,
the difference of
geometric radii, $\Delta R_{\rm geom}$, contains a contribution from
halo effects; hence it is  not  appropriate to define the skin.
The differences $\Delta R_0$ and  $\Delta R_{\rm Helm}$
 both smoothly increase with neutron number, with  $\Delta R_{\rm Helm}$ being
always greater than  $\Delta R_0$, due to the contribution from the
surface thickness. In principle,
both definitions could be used to characterize the skin.
However, due to the smallness of the proton halo (\ref{rhalop}), one
simply has
\begin{equation}\label{rnp}
  \Delta R_{\rm geom}
   \approx
  \delta R_{\rm halo}{\rm(n)}
   +
  \delta R_{\rm skin}
\end{equation}
for
\begin{equation}\label{rskinn}
 \delta R_{\rm skin}\equiv\Delta R_{\rm Helm},
\end{equation}
i.e., Eq.~(\ref{rnp}) gives an additive decomposition into the contributions
to $\Delta R_{\rm geom}$
coming from the weak binding (halo part) and representing
the size effect (skin part)  (see
Fig.~\ref{RadSnall}).
Therefore,
for the present purpose, we prefer $\Delta R_{\rm Helm}$ as a measure
of the skin. Figure~\ref{skin_vdefs}
nicely shows that the neutron halo effect in the Sn
isotopes is predicted to
show up just above $N$=82, and it increases gradually with $N$
reaching in the HFB/SLy4 calculations the value of
$\delta R_{\rm halo}{\rm(n)}$$\approx$0.65$\Delta R_{\rm Helm}$
near the two-neutron drip line.

The results of  calculations for the Ni isotopes
are shown in Fig.~\ref{RadNiall}. Here, the neutron skin quickly
increases above the doubly magic nucleus $^{78}$Ni, i.e.,
above the $N$=50 gap. A simpler pattern is seen for
the Pb isotopes (see Fig.~\ref{RadPbHFB}): the neutron
halo develops for $N$$>$126. In all cases,
$\delta{R}_{\rm halo}$ calculated with SLy4
 is systematically greater than that in  HFB/SkP, RHB/NL3, and RHB/NLSH.

In order to understand these results, we show in
Figs.~\ref{S2nall} and
\ref{S2nPb} the two-neutron separation energies,
$S_{2n}$, for the neutron-rich
Ni, Sn, and Pb isotopes. Systematically, the HFB/SLy4 model
predicts the lowest separation energy.
When approaching the neutron
drip line,  both RHB approaches
yield considerably larger neutron binding than Skyrme-HFB
calculations. This result
is consistent with the model dependence of
$\delta{R}_{\rm halo}$. Indeed the neutron halo
parameter seems to be correlated with the neutron
separation energy. That is, $\delta{R}_{\rm halo}$ increases
with decreasing $S_{2n}$ (see Sec.~\ref{pairing} for more
discussion concerning this point).

The surface  thickness $\sigma$ (\ref{thickness})
shows characteristic
dependence on particle number (see Fig.~\ref{sigmas}). Namely, $\sigma_n$ increases
 with $N$ on the average, but it shows local minima
around magic numbers. This local decrease in $\sigma$ can
be attributed to its sensitivity to pairing correlations
\cite{[Rei96]}. Indeed, static pairing correlations
in magic nuclei vanish, the Fermi surface becomes less diffused,
and the surface thickness is reduced. When approaching the
spherical neutron drip line, $\sigma_n$ behaves fairly smoothly; 
because it is determined from the formfactor at larger $q$
(i.e.,  it seems to be rather
insensitive to the asymptotic behavior of nucleonic density
at large distances). The proton surface  thickness behaves fairly
constant as a function of $N$, although it also
exhibits the local decrease at
magic neutron numbers as a result of self-consistency.

Except for the very neutron-rich nuclei, the RHB models
yield $\sigma$-values which are lower than in the Skyrme-HFB
calculations. This effect is particularly clear for $\sigma_p$,
which is not affected by the variations in the pairing field. In
addition, in all cases $\sigma_p$(NLSH)$<$$\sigma_p$(NL3)
and  $\sigma_p$(SLy4)$<$$\sigma_p$(SkP). (For further discussion, we
refer the reader to Ref.~\cite{[Ben99d]}.)

The difference  $\sigma_n$--$\sigma_p$ exhibits very weak
shell effects. It gradually increases from about 0.2\,fm
around the beta stability line to about 0.5\,fm near
the neutron drip line.
Interestingly, as discussed in Ref.~\cite{[Dob96a]},
the difference between  neutron and proton
radii also depends very weakly on shell effects.

\subsection{Square-well potential analysis}\label{simple}

This section contains some general arguments regarding
the concept of diffraction radius
in two-body halo systems. Our discussion is based on the
spherical finite square-well (SW) potential used in
Ref.~\cite{[Rii92]} to illustrate some  generic aspects
of halos. (See Ref.~\cite{[Mis97]} for the extension to the
deformed case.)

The advantage of this simple model is that by changing the well depth,
one can vary the position of bound single-particle halo orbitals and,
therefore, study the properties of diffraction radii and surface
thickness very close to the $\epsilon$=0 threshold. In our
calculations we assume that the square-well 
potential radius is $R_{\rm SW}$=7\,fm and
the system consists of 70 particles.  The potential depth is varied
to tune the energy of the
last bound nucleon. The halo structure is represented by two neutrons
in the $3s_{1/2}$ orbital, while the core can be associated with the
remaining 68 particles occupying well-bound states.

With the binding energy of the $3s_{1/2}$  orbital approaching
zero, the halo develops. This is illustrated in
Fig.~\ref{densities_sqw}, which shows the total
and core  densities
for three values of  the binding energy
of the $3s_{1/2}$ halo orbital: --5\,MeV,
--100\,keV, and --1\,keV. 
The
presence of the halo  is clearly seen at larger distances,  $r$$>$8\,fm.

The
form factors of the total  and core
 densities at $\epsilon$($3s_{1/2}$)=--5\,MeV
and --1\,keV are displayed in
Fig.~\ref{F_sqw}.
As a consequence of the uncertainty principle,
in the case of a very weak binding
the  form factor of the halo
wave function  (shown in the insert) corresponds
to a very narrow
momentum distribution, and it contributes very little to the
total form factor. Consequently, the
first zero of the form factor, hence the
diffraction radius, is very weakly influenced by the
presence of the halo. This is not true in the case of a
large binding where the valence orbital does not have the
halo character,  and its form factor is significantly greater
from zero in the region of $q_1$.

Figure~\ref{radii_sqw} displays the
radii calculated
as a function of the $3s_{1/2}$ binding energy.
Due to the halo character of the valence orbital,
with the energy of the $3s_{1/2}$ state approaching zero,
the total  geometric  radius diverges as $(-\epsilon)^{-1}$
\cite{[Rii92]}. At the same time, the geometric radius of the core,
as well as the Helm radii for the total system
($R_{\rm Helm,total}$)
and the core ($R_{\rm Helm,core}$),
very weakly depend on $\epsilon$.
The  effect of decreased binding on the core is  measured
by  the difference $R_{\rm geom,core}$$-$$R_{\rm Helm,core}$,
which gives
the core  contribution to $\delta{R}_{\rm halo}$.
As expected,
in the limit of weak binding, the halo
parameter is almost entirely determined by the
asymptotic behavior of the $3s_{1/2}$ wave function. It
is also seen that  the
difference between $R_{\rm Helm,total}$ and
 $R_{\rm Helm,core}$
is very small.

\subsection{Pairing anti-halo effect}\label{pairing}

In contrast to light nuclei where the halo can be associated
with very few weakly bound neutrons that are practically
decoupled from the rest of the system, it is difficult to separate
halo structures in heavier systems
in which {\em all} the nucleons (including
the valence weakly bound  neutrons) move in
one self-consistent field.
An additional difference and complication
is caused  by the presence of strong pairing
correlations in heavy open-shell nuclei. As found in Ref.~\cite{[Ben00]}
and discussed below,
pairing  strongly modifies the extreme single-particle picture
of halo structures
presented in Sec.~\ref{simple}.

Consider, e.g., the valence neutron moving in a
 mean-field
 potential.
Due to the fact that the nuclear mean field vanishes at large distances,
the standard asymptotic
behavior of the neutron density is, in the absence of pairing correlations,
given by
\begin{equation}\label{as1}
\rho(r) \rightarrow \sim \frac{\exp{(-2\kappa r)}}{r^2},
\end{equation}
where
\begin{equation}
\kappa =\sqrt{\frac{2m(-\epsilon)}{\hbar^2}},
\end{equation}
with $\epsilon$ being the single-particle energy of the least bound
neutron.
In the presence of pairing,
the  constant $\kappa$ is different \cite{[Bul80],[Dob84],[Dob96],[Ben00]},
and for the even neutron numbers reads
\begin{equation}\label{chi}
\kappa =\sqrt{\frac{2m(E_{\rm min}-\lambda)}{\hbar^2}},
\end{equation}
where $E_{\rm min}$ is the lowest quasi-particle
energy and $\lambda$ is the Fermi energy.

In the extreme single-particle picture (no pairing), the
halo structure may develop when $\epsilon \rightarrow 0$
($\kappa \rightarrow 0 $) \cite{[Rii92]}.
However, according to
expression (\ref{chi}), in the limit of vanishing binding
($\lambda \rightarrow 0$) the  constant  $\kappa$ does not vanish and reads
\begin{equation}\label{chimin}
\kappa_{\rm min} \approx  \sqrt{\frac{2m\Delta}{\hbar^2}},
\end{equation}
where $\Delta$ is the pairing gap of the lowest
quasiparticle. Consequently, in the presence of pairing correlations,
$\kappa$ is never small, and a huge halo, as it is seen in light nuclei,
cannot develop ({\em{pairing anti-halo effect}} of Ref. \cite{[Ben00]}).

In order to confirm the influence of pairing and weak binding
on  $\delta R_{\rm halo}$, we performed spherical HFB/SLy4
calculations near the two-neutron and two-proton drip lines.
Figure~\ref{boundsn}
shows the neutron halo parameters, neutron Fermi energies, and
 neutron pairing gaps  calculated in the
HFB/SLy4  model for the two-neutron drip-line even-even nuclei
(which are the heaviest  even-even isotopes that are still
predicted to be two-neutron bound). First, we note that $\Delta_n$
does not vanish near the two-neutron drip line. This phenomenon
has been found and discussed in detail in
Refs.~\cite{[Dob84],[Bel87],[Dob96],[Smo93]}, and it
was attributed to the strong coupling to the neutron continuum
in the pairing channel.
Second, the pairing gap shows some shell fluctuations:
 the minima in  $\Delta_n$ appear
at $N$=86, 130, and 192, i.e., just above
the neutron magic gaps. On the average, however,
 $\Delta_n$ stays between $\sim$1.8\,MeV in light nuclei and $\sim$1.2\,MeV
in the heaviest elements. As a result, the exponent
(\ref{chimin}) is always
 sizable, and $\delta R_{\rm halo}$(n) does not exceed 1\,fm in heavy
even-even nuclei.

The pattern of  $\delta R_{\rm halo}$(n) seen in Fig.~\ref{boundsn}
is  nicely correlated with the behavior of $\lambda_n$. Namely, the
neutron halo parameter increases when the Fermi energy approaches zero.
It is to be noted, however, that there is no clear correlation between
the magnitude of  $\delta R_{\rm halo}$(n) and the appearance of
low-$\ell$ ($s$ and $p$) states at the Fermi energy \cite{[Rii92]}, as one
would expect from an extreme single-particle picture
of Sec.~\ref{simple}. It seems that the
pairing anti-halo
effect is far more important than the influence of the centrifugal barrier,
cf.~discussion in Ref. \cite{[Ben00]}.
We made an attempt to find a phenomenological expression that
would express
$\delta R_{\rm halo}$(n)
in terms of $(\xi \Delta_n-\lambda_n)^\eta$ ($\xi, \eta$
being  free
parameters). Unfortunately, we were not able to obtain
a unique fit for all neutron-weak nuclei at once,
although  some  correlation between these two quantities
exists.

The insert in Fig.~\ref{boundsn}  shows the proton halo parameter
in the least bound even-even isotones near the two-proton drip line.
As expected, due to the
confining effect of the Coulomb potential, the proton halo is very
small -- of the order of 0.02\,fm. It is only in the very light
$sd$ nuclei that $\delta R_{\rm halo}$(p) can exceed 0.1\,fm.
Interestingly, there is also some increase in the proton halo in
the superheavy nuclei with $Z$$\sim$120, $N$$\sim$172, which, in
some spherical calculations, show bubble-like structures
\cite{[Dec99],[Ben99a]}. In this context, it should be
emphasized again that calculations shown in Fig.~\ref{boundsn}
are spherical, and some modifications due to deformation are
expected; in particular, the superheavy nucleus with $Z$=120 and
$N$=172 is not expected to be spherical in the HFB/SLy4 model
\cite{[Cwi96],[Bur98],[Cwi99]}.

\subsection{Global behavior of halos and skins in spherical even-even nuclei}

In order to study  the systematic behavior
of the spherical-shape  density distributions, we performed systematic
calculations
in the spherical HFB/SLy4 and HFB/SkP models for all
even-even nuclei
predicted to be stable with respect to the two-nucleon emission, i.e.,
for all even-even nuclei with positive two-neutron and two-proton
separation energies,
$S_{2n}$=$B(N,Z)-B(N$$-$$2,Z)$$>$0 and
$S_{2p}$=$B(N,Z)-B(N,Z$$-$$2)$$>$0.

The results for the  neutron halos are shown in
Figs.~\ref{halon_SLy4} and \ref{halon_SkP} for the HFB/SLy4 and
HFB/SkP models, respectively.
Several features seen in
these systematics  are noteworthy. Firstly,  for most nuclei
$\delta R_{\rm halo}$(n) is very small.
Only in the immediate vicinity of the two-neutron drip line is
a rapid increase in the halo parameter seen. As discussed above,
while the halo effect is rather strong for
the SLy4 force, the HFB/SkP model predicts very
few candidates for a halo.

A second interesting aspect is a weak dependence of the halo
parameter on shell effects.  Contrary to the rms radii which
show a significant reduction  around spherical magic gaps
\cite{[Dob96a]}, the variations of $\delta R_{\rm halo}$(n) around
magic gaps are much weaker. This is easy to understand. In
well-bound nuclei where the shell effects are very pronounced,
the halo parameter is dramatically reduced due to
weak binding. On the other hand,
in neutron drip-line nuclei,  shell effects are 
 significantly weakened (reduced
magic gaps, strong  pairing correlations); hence their influence
on radii is less significant. The pattern shown in
Figs.~\ref{halon_SLy4} and \ref{halon_SkP} basically reflects
the behavior of the two-neutron separation  energy around the
two-neutron drip line, and is qualitatively similar to that
for $\Delta R_{\rm np}$ discussed in Ref.~\cite{[Dob96a]}.

As shown in Fig.~\ref{boundsn}, the proton halo parameter is
much smaller than  that for the neutron halo. The inset in
Fig.~\ref{halon_SLy4} shows $\delta R_{\rm halo}$(p) for very
light nuclei. The largest proton halos, $\delta R_{\rm
halo}$(p)$\sim$0.15\,fm,  can be found around $^{20}$Mg and
$^{24}$Si.  For more discussion, see Sec.~\ref{sec:3D}.

Figure~\ref{skins_SLy4} shows the neutron skins calculated in
the HFB/SLy4 model. One sees that the skin grows steadily in a
direction orthogonal to the valley of stability.
The weak mass dependence and a nearly linear trend with
the neutron excess $N$$-$$Z$ suggests that $\delta R_{\rm
skin}$ reflects the bulk size properties of neutrons and protons.
As discussed in Ref.~\cite{[Dob99a]},
the isovector dependence of the neutron skin is
governed  by a balance between the volume \cite{[Ben99d],[Rei99d]}
and surface symmetry energy coefficients. Within the present sample, this is
confirmed by the fact that the HFB/SkP results for the neutron
skin
are indeed very similar, and it is noted that SkP and SLy4 do have a
very similar symmetry energy coefficient (32\,MeV in SLy4
and 30\,MeV in SkP).

Last but not least, it is worth inspecting  the global trends of
the surface thickness $\sigma_{\rm p,n}$. Figure~\ref{sigman}
shows the neutron surface thickness. 
It s pattern shares one feature with the neutron halo (Fig.~\ref{halon_SLy4}), namely 
$\sigma_{\rm n}$ is
particularly large near the neutron drip line.
However, $\sigma_{\rm n}$  displays a much richer
 structure all over the periodic table with maxima far from
 closed shells.
 The proton surface thickness is
shown in Fig.~\ref{sigmap}. Compared to $\sigma_n$,  the global
behavior of $\sigma_p$ is  different.  The variations with
$N$$-$$Z$  are less systematic and, in some cases, $\sigma_p$ {\em
decreases} when approaching the two-proton drip line. As in
neutrons, the proton surface thickness is reduced around magic
gaps.  As discussed earlier, $\sigma_{\rm p}$ is generally
much smaller than $\sigma_{\rm n}$.

It is  interesting to note the presence of an island of
particularly small neutron and proton skins near the proton drip
line in the region of superheavy nuclei  with  $N$$\sim$172.
This is probably related to the pronounced dip of the spherical
distribution near the nuclear center which appears for these
nuclei \cite{[Dec99],[Ben99a]}. For the protons, there exists a
further island of small surface thickness for superheavy
elements with $N$$>$200.  However, as discussed in
Sec.~\ref{pairing}, the presence of bubble-like structures in
this region may be an artifact of the assumption of spherical
symmetry \cite{[Cwi99b]}.

Finally, the neutron-proton difference of the surface thickness
is shown in Fig.~\ref{deltasigmas}. It displays a mix of steady
growth with $A$ as well as with $N-Z$ which is just the sum of
the different trends seen for protons and neutrons separately.
Like  the diffraction radii and rms (or geometric) radii, the
shell effects which are present in both observables are somewhat
suppressed in the differential quantity.

\subsection{RHB calculations of halos in the Ne isotopes}

In contrast to the simple model of Sec.~\ref{simple}
or analysis  of Ref.~\cite{[Ben00]},
in microscopic calculations the binding energy of a single
particle (or quasi-particle) orbital is not a free parameter
but is obtained self-consistently from the realistic
Hamiltonian. Hence it is difficult to find a case where
a low-$\ell$ orbital (i.e., a potential candidate for halo) appears
very close to the threshold.
Here we discuss the case studied in Ref.~\cite{[Poe97a]}, where,
based on the RHB/NL3 model, such a situation was found for
the neutron-rich Ne isotopes. According to this  work,
the neutron Fermi energy in the Ne isotopes with $N$$>$20
stays very close to zero, stabilized by the presence of
 three close-lying single-particle
canonical orbitals,
$2p_{3/2}$, $2p_{1/2}$, and $1f_{7/2}$.

The low-$\ell$ shells,
$2p_{3/2}$ and $2p_{1/2}$,  are good candidates for halo orbitals. It is worth
noting that their canonical  HFB energies stay very close in energy. This
suggests that the wave functions of the $2p_{3/2}$ and $2p_{1/2}$ are weakly
influenced by the spin-orbit interaction
 (a situation that is characteristic of halo states)
\cite{Lala98}.
This is nicely
demonstrated in Fig.~\ref{ffwfall} which shows the form factors of single-neutron
canonical states $1p_{3/2}$, $1d_{3/2}$,  $2p_{3/2}$,  and $2p_{1/2}$
in the drip-line nucleus $^{38}$Ne. The form factors of the $2p_{3/2}$ and $2p_{1/2}$
orbitals  are  very similar which confirms the
 negligible effect of the
spin-orbit interaction on these states. The narrow momentum distribution of the
$2p$ orbitals
is indicative of weak binding. In contrast, wider form factors of
 well-bound
$1p_{3/2}$ and  $1d_{3/2}$ orbitals reflect the fact that their
wave functions are better localized inside the nuclear volume.

The neutron distribution form factors  in $^{20,34}$Ne obtained in RHB/NL3
are shown in Fig.~\ref{ff2034}. In $^{20}$Ne,  the contribution from the  valence
 $1d_{5/2}$ neutrons has been singled out, and it is seen that the influence
 of the valence orbits on the diffraction radius is strong. On the other hand,
the effect of the $2p$ valence orbitals on $R_0$ in  $^{34}$Ne is small. The reason
for this is twofold. Firstly, in accordance with the discussion from Sec.~\ref{simple},
the $2p$ form factor is narrow and it mainly contributes
around $q$=0. Secondly, there are less than  two neutrons in the $2p$ shell. Hence
the behavior of the total form factor  in $^{34}$Ne (the ratio of the number of neutrons
in valence orbits to that in the core
 is $\sim$0.08) is primarily governed by the core neutrons.
(In $^{20}$Ne the valence/core ratio is 0.25.)

The above situation discussed for $^{34}$Ne is, in fact, typical for all
weakly bound neutron-rich nuclei. The  weak binding of valence  orbits and the fact
that they are occupied by very few particles makes the diffraction radius weakly
dependent on halo structures. It is interesting to see that the simple argument presented
in Sec.~\ref{simple} works in a microscopic case, in spite of the fact that  pairing
modifies the naive single-particle picture to some extent.

\subsection{Charge halo parameter in stable nuclei}
\label{sec:3D}

Elastic electron scattering has provided a world of well-evaluated
data on nuclear charge distributions; see, e.g., Ref.~\cite{[Fri82],[DeV87]}.
It is nicely corroborated by the very precisely measured
root-mean-square charge radii \cite{[Fri95]}. This offers a possibility
to deduce the experimental charge halo parameter,
\begin{equation}\label{rch}
\delta R_{\rm halo}{\rm(ch)} \equiv R_{\rm geom}{\rm(ch)} - R_{\rm Helm}{\rm(ch)},
\end{equation}
in selected cases.  In order to have a most complete and up-to-date
supply of data on  charge radii and surface thickness, we
have recurred to the data base of  Ref.~\cite{[Fri99]}, which is a compilation
of results analyzed as explained in \cite{[Fri82]}.

The charge form factor $F_{\rm C}(q)$ is composed of the
form factors of the proton and neutron distribution multiplied
by the corresponding
 intrinsic nucleon form factors.  A similar
contribution from the magnetic form factors is
added. Finally, the  centre-of-mass
correction 
is performed.
 The nucleon form factors are taken from electron
scattering on the proton and the deuteron \cite{simon,walther}
and parametrized in terms of the Sachs form factors as outlined
in Ref.~\cite{FriNeg}. A detailed description 
of the procedure used to determine the charge radii
can be found in
Appendix 2 of \cite{RMFrev}.  From the charge form factor,
 we deduce the rms radius, the diffraction
radius, and the surface thickness in the standard manner.

 Figure~\ref{halo_ch}
 shows the experimental  values of $\delta R_{\rm halo}$(ch)
 for selected nuclei (doubly magic $^{16}$O, $^{40,48}$Ca, $^{58}$Ni,
 $^{208}$Pb, semi-magic $^{52}$Cr, $^{54}$Fe, $^{88}$Sr,  $^{90}$Zr, $^{92}$Mo,
 $^{116,124}$Sn,  $^{204,206}$Pb, and some open-shell nuclei, including
 the well-deformed Cr and Sm isotopes). They are displayed
 together with predictions of spherical
Skyrme HFB calculations.

It is to be noted that the charge halo
 is a very
sensitive observable because it stems from subtracting two large
radii (\ref{rch}).  The experimental error on  $\delta R_{\rm halo}$(ch) is
at least as large as the
largest error on radii and surface thickness. This leads to a conservative
 uncertainty on the data, $\sim$0.03\,fm.
As  expected from our  results on proton halos, the charge halos
are all very small. A notable exception is $^{16}$O where
$\delta R_{\rm halo}$(ch) is 0.13\,fm. Also, our calculations
are expected to slightly underestimate charge halos in some
open-shell nuclei (e.g., $^{152,154}$Sm) due to possible contributions from
deformation effects.  Considering the above,  it
is very satisfying to see that
our HFB results are generally close to the experimental points, in
fact staying within the experimental uncertainty in most cases.
Since proton halos  are close to the
charge halos  in all calculations, this nice
agreement, together with the systematic behavior of proton halos discussed
above, make us conclude that the pronounced proton halos
do not exist.

At second glance, one is tempted to spot shell effects and isotopic
trends when looking at the fluctuation of the charge halos in
Fig.~\ref{halo_ch}. However,  these variations
stay within the experimental uncertainties and cannot serve for
a deeper
analysis. The charge halos as such are too small and, moreover, the
regime of stable nuclei does not supply enough variation for that.

\section{Conclusions}\label{conclusions}

This work contains the theoretical analysis of neutron and
proton skins, halos, and surface thickness obtained within the
spherical self-consistent mean-field theory.  The main goal was
to describe spatial characteristics of nucleonic densities of
nuclei far from stability, where the closeness of the particle
continuum qualitatively changes the physical situation.

The Helm model analysis presented in this work allows for
an unambiguous determination of halos and skins from
nucleonic density distributions. It has been  shown that the halo
parameter, defined as  the difference between the geometric radius
(a rescaled rms radius)
and the Helm radius, is small in well-bound nuclei, and for neutrons it
becomes enhanced for heavy exotic systems with low neutron
separation energies. However, unlike in the  neutron-rich
few-body systems, our
calculations do not predict giant neutron halos in medium-mass
and heavy nuclei. This is because strong pairing correlations
effectively reduce the impact of weak binding on
the asymptotic behavior of the single-particle density
(pairing anti-halo effect \cite{[Ben00]}).

No significant  proton halo  has been found
when approaching the proton drip line. A moderate
effect (less than 0.2\,fm) is predicted for some light nuclei,
but it can be practically neglected for heavier systems.
The experimental values of charge halos for stable nuclei,
of the order of 0.02-0.04\,fm, are perfectly consistent with the
mean-field predictions.

The neutron skin, defined as a difference of neutron and proton
Helm radii, shows a smooth gradual dependence on the neutron
excess and  is extremely weakly affected by shell effects. This
is consistent with the results of a previous study \cite{[Dob96a]}
where a very weak shell dependence of
$\Delta R_{np}$ was found.

On average, the neutron surface thickness increases
with neutron number, but it is
locally reduced around magic numbers,
thanks to reduced pairing. On the other hand,
proton surface thickness depends to a lesser degree on proton
number; it rather
tends to follow the trend dictated by $\sigma_n$.
As a result, the  difference $\sigma_n-\sigma_p$ shows
a reduced dependence on shell effects. A very interesting
situation is predicted for the superheavy $N$=172 isotones
where the proton surface thickness is actually reduced with
increasing proton number.

Theoretically,
the analysis based on density form factors is very simple
and physically elegant.
Unfortunately, experimental information on nucleonic densities
is currently limited to charge densities in some stable nuclei,
and almost nothing is known on neutron density distributions.
Starodubsky and Hintz \cite{[Sta94]} made an attempt to deduce
neutron densities in $^{206,207,208}$Pb -- in a model-dependent way --
from  elastic proton scattering
at intermediate energies, and this marks the
 state-of-the-art. An exciting new avenue is a prospect
for  {\em direct}
measurements of the neutron density form factors from
the asymmetry in the parity-violating elastic polarized
electron scattering
\cite{[Don89],[Don92],[Pol92],[Hor93],[Alb99],[Vret99]}.
Table~\ref{table1} shows the Helm model
analysis of  neutron densities
in $^{40}$Ca and $^{208}$Pb calculated in our
self-consistent models. 
Theoretical predictions for diffraction radii are rather
robust, with
the differences between
values of $R_0$ 
obtained in  different models being  below 0.5\% for 
$^{40}$Ca and below 3.5\% for  $^{208}$Pb. Interestingly,
due to  a compensation 
effect between $R_0$ and $\sigma$,
the geometric neutron radius in $^{208}$Pb
is rather similar in all models,
 $R_{\rm geom}$$\approx$7.32\,fm.
We hope that our results will stimulate future experimental
studies of neutron distributions in nuclei.

\acknowledgments

This research was supported in part by the U.S. Department of
Energy under Contract Nos.\ DE-FG02-96ER40963
(University
of Tennessee), DE-FG05-87ER40361 (Joint Institute for Heavy
Ion Research), DE-AC05-96OR22464 with Lockheed Martin Energy
Research Corp. (Oak Ridge National Laboratory),
by the Polish Committee for Scientific
Research (KBN) under Contract No.~2~P03B~040~14, and
NATO grant CRG 970196.

%\bibliography{wn1,wnhalo}
%\bibliographystyle{unsrt}

\bigskip
\begin{table}
\caption{Characteristics of 
neutron  distributions  in 
$^{40}$Ca and $^{208}$Pb: diffraction radius $R_0$,
surface thickness $\sigma$, 
and geometric radius $R_{\rm geom}$  (all in fm),
obtained in the HFB and RMF models employed in this
work. 
}
\begin{tabular}{llcccc}
  nucleus  &  & SLy4 & SkP  & NL3 & NLSH  \\
\hline
 & $R_0$ & 3.827 & 3.844 & 3.844 & 3.841 \\   
 $^{40}$Ca &$\sigma$ & 0.905 & 0.923 & 0.845 & 0.793 \\ 
 & $R_{\rm geom}$ & 4.353  & 4.388  & 4.296 & 4.274 \\
\hline
 & $R_0$ & 6.870 & 6.849 & 7.076 & 7.075 \\   
 $^{208}$Pb  & $\sigma_n$ & 1.022 & 1.033 & 0.971 & 0.929 \\ 
 & $R_{\rm geom}$ & 7.252  & 7.244  & 7.409 & 7.374 
\end{tabular}
\label{table1}
\end{table}
\clearpage

\newcommand{\myclearpage}{\clearpage}
\renewcommand{\myclearpage}{}

\newcommand{\myclearpagew}{\clearpage}

\newlength{\myplotsize}
\setlength{\myplotsize}{8cm}
\newlength{\myplotsizew}
\setlength{\myplotsizew}{16cm}

\newlength{\myplotsizen}
\setlength{\myplotsizen}{13cm}

\begin{figure}
\vbox{
\begin{center}
\leavevmode
\epsfxsize=\myplotsize
\epsfbox{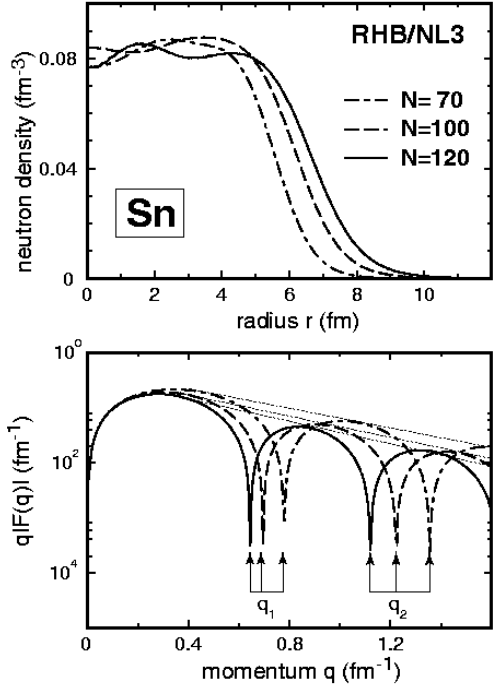}
\end{center}
\caption{Top: neutron densities calculated in the RHB/NL3
 model for $^{120,150,170}$Sn. Bottom: the
corresponding form factors. Positions of the first and
second zeros in the form factors are indicated by arrows.
}
\label{rhoF(Sn)}
}
\end{figure}
\myclearpage

\begin{figure}
\vbox{
\begin{center}
\leavevmode
\epsfxsize=\myplotsize
\epsfbox{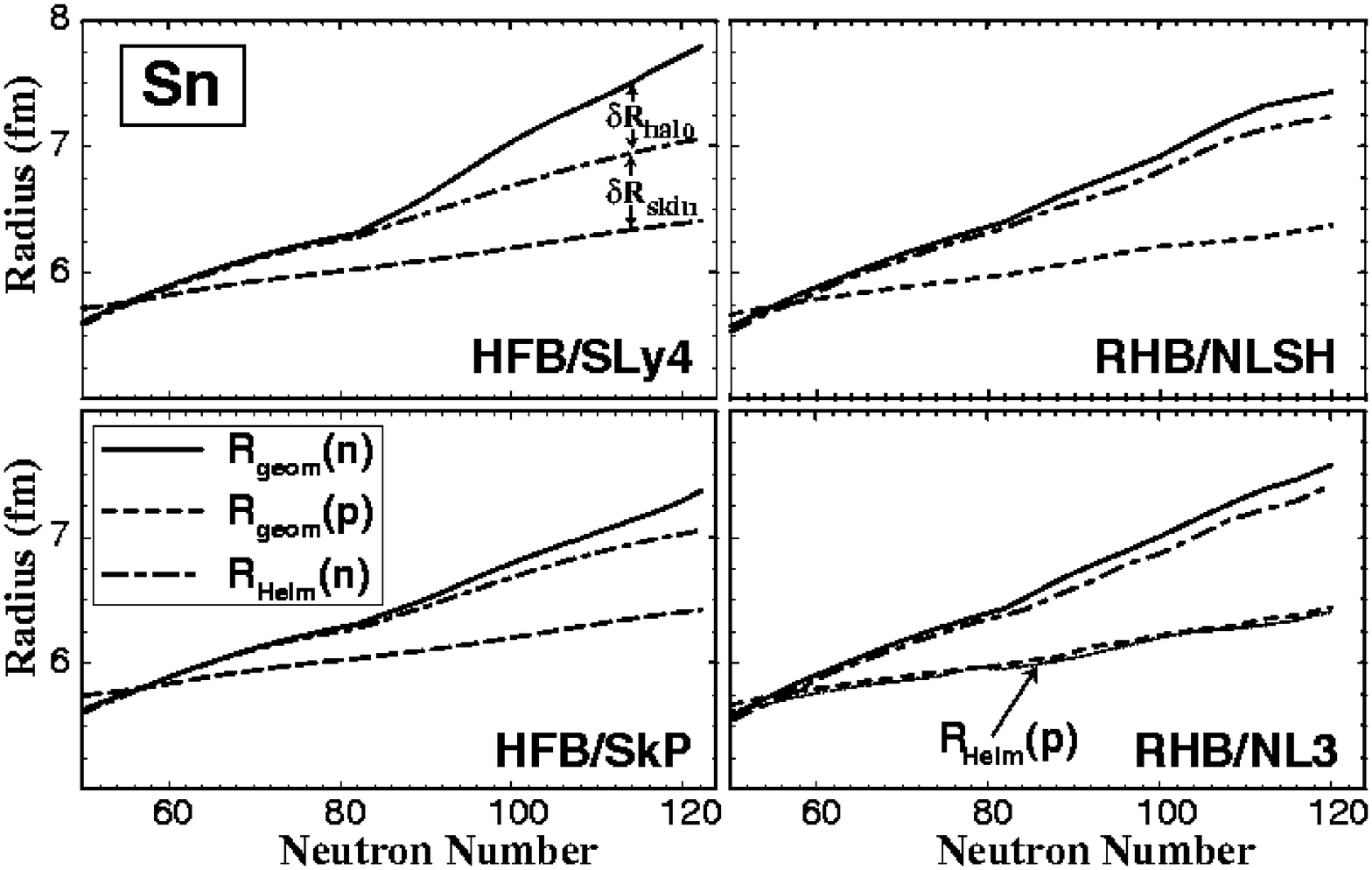}
\end{center}
\caption{Neutron  $R_{\rm geom}$(n) and
$R_{\rm Helm}$(n),  and proton $R_{\rm geom}$(p)
radii for the Sn isotopes
calculated in the HFB/SLy4, HFB/SkP,
RHB/NLSH, and RHB/NL3
models. The proton Helm radius $R_{\rm Helm}$(p) is also shown
in the  RHB/NL3 variant (dotted line); it is very close to
$R_{\rm geom}$(p).
}
\label{RadSnall}
}
\end{figure}
\myclearpage

\begin{figure}
\vbox{
\begin{center}
\leavevmode
\epsfxsize=\myplotsize
\epsfbox{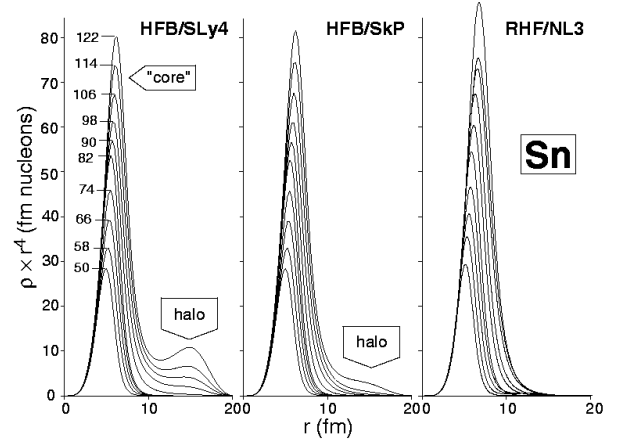}
\end{center}
\caption{Neutron densities multiplied by $r^4$ for the tin isotopes
with $N$=50, 28, 66, 74, 82, 90, 98, 106, 114, and 122
calculated in the HFB/SLy4, HFB/SkP, and RHB/NL3 models.
}
\label{rhor4}
}
\end{figure}
\myclearpage

\begin{figure}
\vbox{
\begin{center}
\leavevmode
\epsfxsize=\myplotsize
\epsfbox{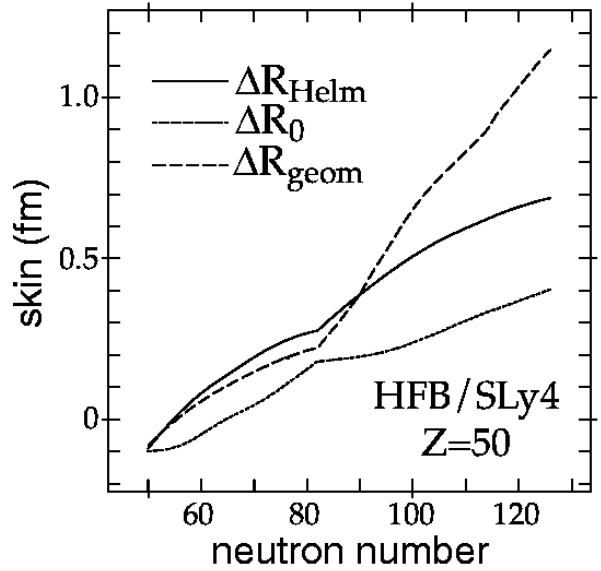}
\end{center}
\caption{The neutron-proton radius differences (\protect\ref{rskin})
for the even-even Sn isotopes
calculated in the HFB/SLy4
model.
}
\label{skin_vdefs}
}
\end{figure}
\myclearpage

\begin{figure}
\vbox{
\begin{center}
\leavevmode
\epsfxsize=\myplotsize
\epsfbox{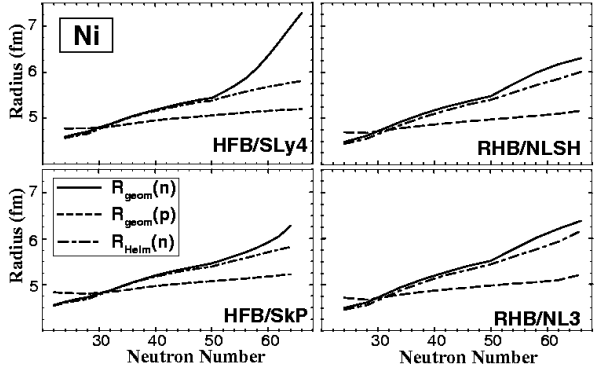}
\end{center}
\caption{Same as in Fig.~\protect\ref{RadSnall} except
 for the Ni isotopes.
}
\label{RadNiall}
}
\end{figure}
\myclearpage

\begin{figure}
\vbox{
\begin{center}
\leavevmode
\epsfxsize=\myplotsize
\epsfbox{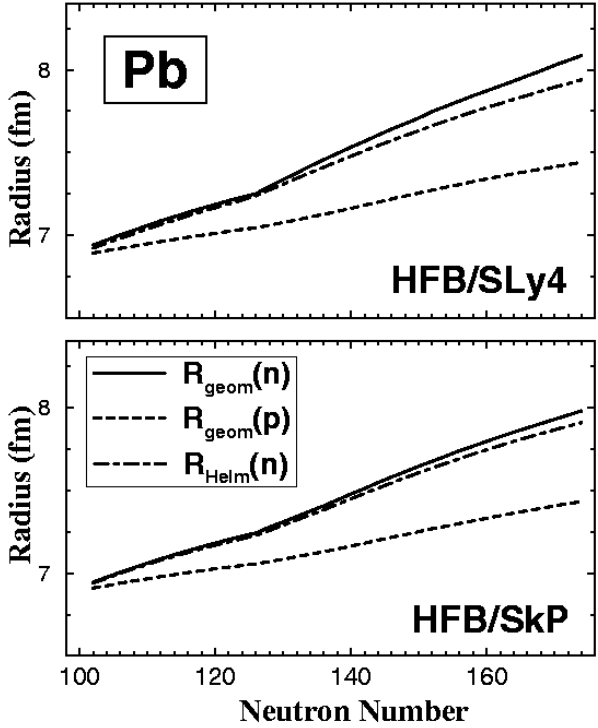}
\end{center}
\caption{Neutron  and
proton
radii for the Pb isotopes
calculated in the HFB/SLy4 and HFB/SkP
models.
}
\label{RadPbHFB}
}
\end{figure}
\myclearpage

\begin{figure}
\vbox{
\begin{center}
\leavevmode
\epsfxsize=\myplotsize
\epsfbox{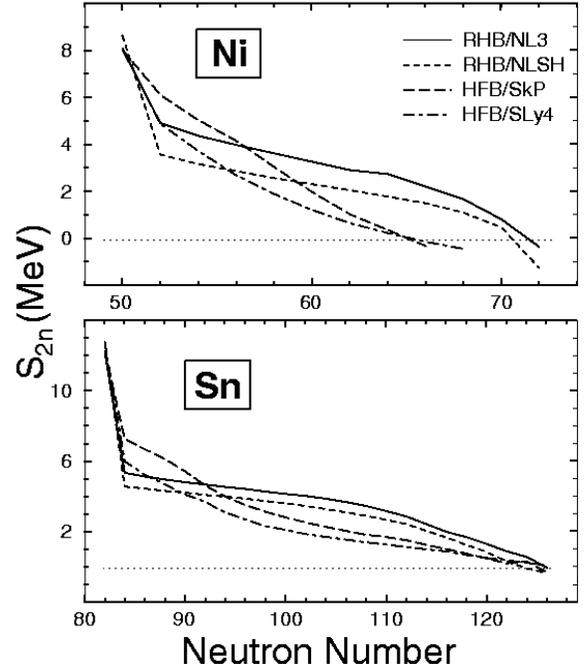}
\end{center}
\caption{Two-neutron separation energies for the neutron-rich
Ni (top) and Sn (bottom)
isotopes calculated in the HFB/SLy4, HFB/SkP,
RHB/NLSH, and RHB/NL3
models.
}
\label{S2nall}
}
\end{figure}
\myclearpage

\begin{figure}
\vbox{
\begin{center}
\leavevmode
\epsfxsize=\myplotsize
\epsfbox{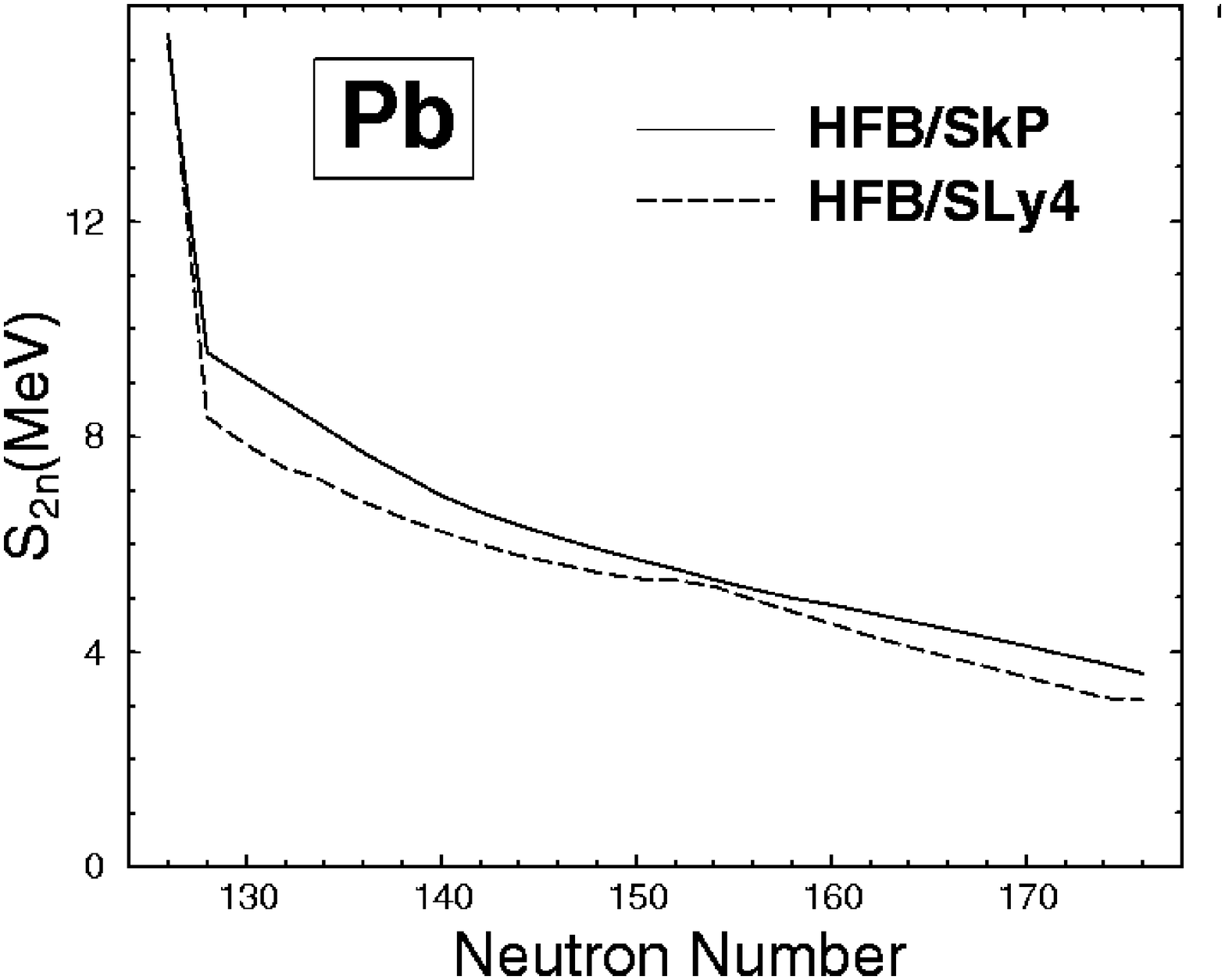}
\end{center}
\caption{Two-neutron separation energies for the neutron-rich
Pb
isotopes calculated in the HFB/SLy4 and HFB/SkP
models.
}
\label{S2nPb}
}
\end{figure}
\myclearpage

\begin{figure}
\vbox{
\begin{center}
\leavevmode
\epsfxsize=\myplotsize
\epsfbox{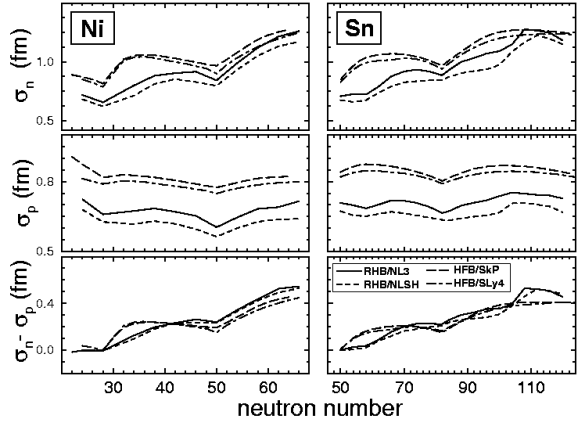}
\end{center}
\caption{Neutron (top) and proton (middle)
surface thickness coefficients for the Ni (left)
and Sn (right) isotopes calculated in the
RHB/NL3, RHB/NLSH, HFB/SkP, and HFB/SLy4 models.
The difference, $\sigma_n$$-$$\sigma_p$,
is shown in the bottom panels.
}
\label{sigmas}
}
\end{figure}
\myclearpage

\begin{figure}
\vbox{
\begin{center}
\leavevmode
\epsfxsize=\myplotsize
\epsfbox{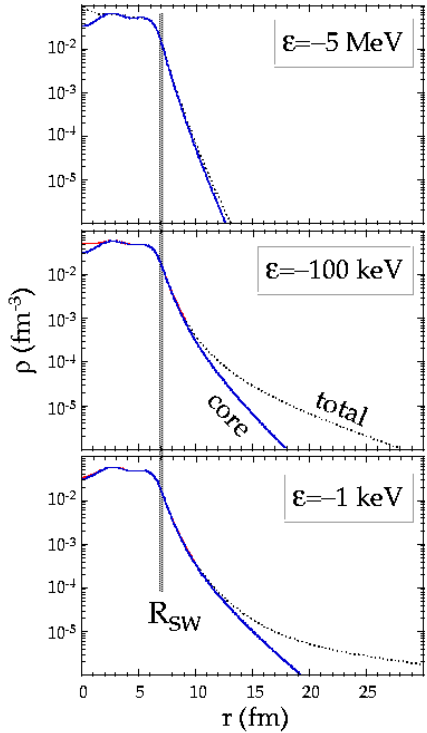}
\end{center}
\caption{Total (dotted line) and core
(solid line) densities for the system of 70 particles
moving in the 
the finite spherical
square-well potential with the radius $R_{\rm SW}$=7\,fm for
the three
values of the binding energy
of the $3s_{1/2}$ halo orbital: --5\,MeV (top),
--100\,keV (middle), and --1\,keV (bottom).
}
\label{densities_sqw}
}
\end{figure}
\myclearpage

\begin{figure}
\vbox{
\begin{center}
\leavevmode
\epsfxsize=\myplotsize
\epsfbox{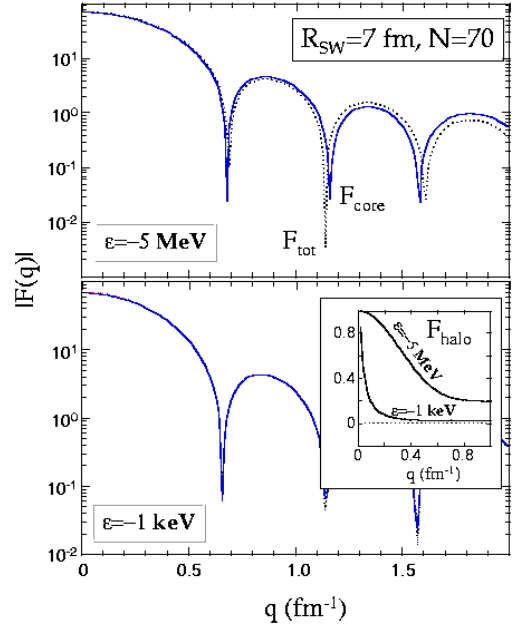}
\end{center}
\vspace{-1cm}
\caption{Form factors of the total (dotted line) and core
(solid line) densities for the system of 70 particles
moving in the finite spherical
square-well potential with the radius $R_{\rm SW}$=7\,fm for
the two
values of the binding energy
of the $3s_{1/2}$ halo orbital: $-$5\,MeV (top)
and $-$1\,keV (bottom). The form factor of the halo
wave function is shown in the inset.
}
\label{F_sqw}
}
\end{figure}
\myclearpage

\begin{figure}
\vbox{
\begin{center}
\leavevmode
\epsfxsize=\myplotsize
\epsfbox{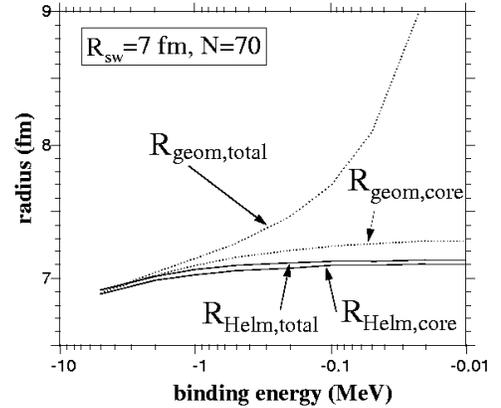}
\end{center}
\caption{Dependence of the geometric radii, namely
the total, $R_{\rm geom,total}$, and core,
$R_{\rm  geom,core}$ radii and of the corresponding Helm radii,
$R_{\rm Helm,total}$ and $R_{\rm Helm,core}$, on the binding energy
of the $3s_{1/2}$ halo orbital of the finite square
well with radius $R_{\rm SW}$=7\,fm. The total number of particles
is $N$=70. The core consists of 68 particles occupying
all the single-particle orbitals below $3s_{1/2}$.
}
\label{radii_sqw}
}
\end{figure}
\myclearpagew
\newpage

\widetext
\begin{figure}
\vbox{
\begin{center}
\leavevmode
\epsfxsize=\myplotsizen
\epsfbox{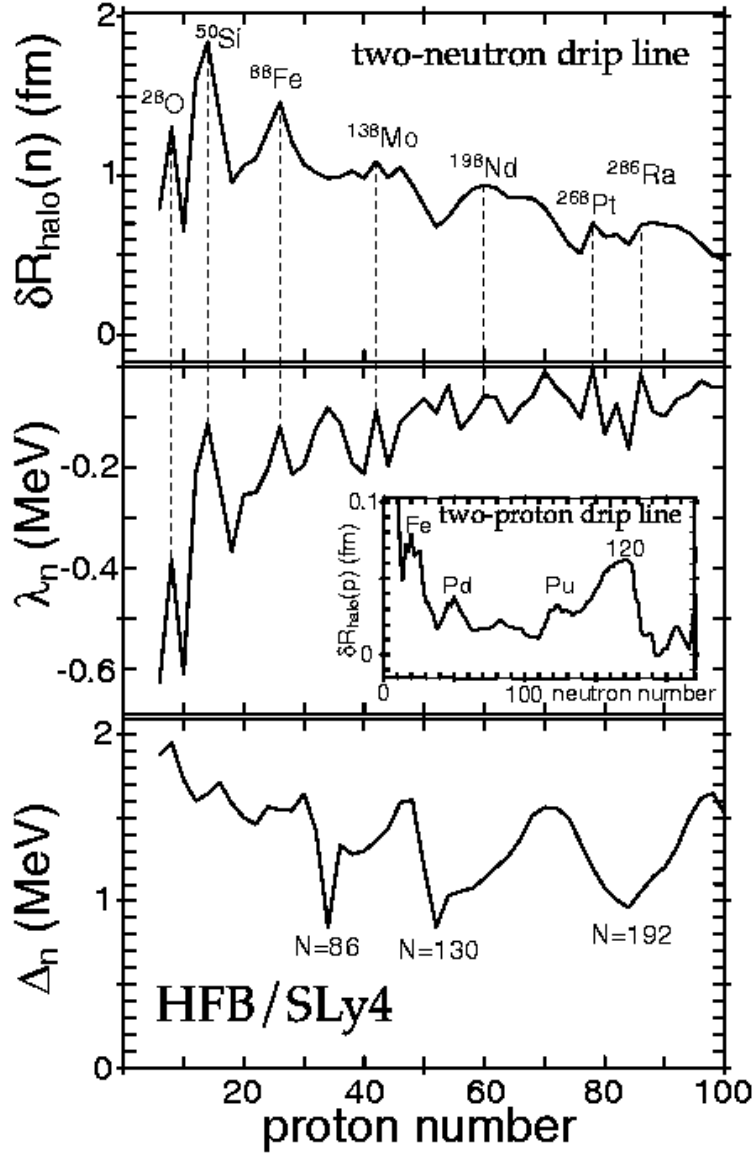}
\end{center}
\caption{Neutron halo parameters (top), neutron Fermi energies (middle),
and neutron pairing gaps (bottom) calculated in the
HFB/SLy4  model for the two-neutron drip-line even-even nuclei
(i.e., the heaviest  even-even isotopes which are
predicted to be two-neutron bound).  The proton halo parameters for
the two-proton drip-line even-even nuclei are shown in the inset.
Note the correlation between $\delta R_{\rm halo}$
and $\lambda_n$, marked by the vertical dashed lines; whenever $\lambda_n$
approaches zero, $\delta R_{\rm halo}$ tends to increase.
}
\label{boundsn}
}
\end{figure}
\myclearpagew

\widetext
\begin{figure}
\vbox{
\begin{center}
\leavevmode
\epsfxsize=\myplotsizew
\epsfbox{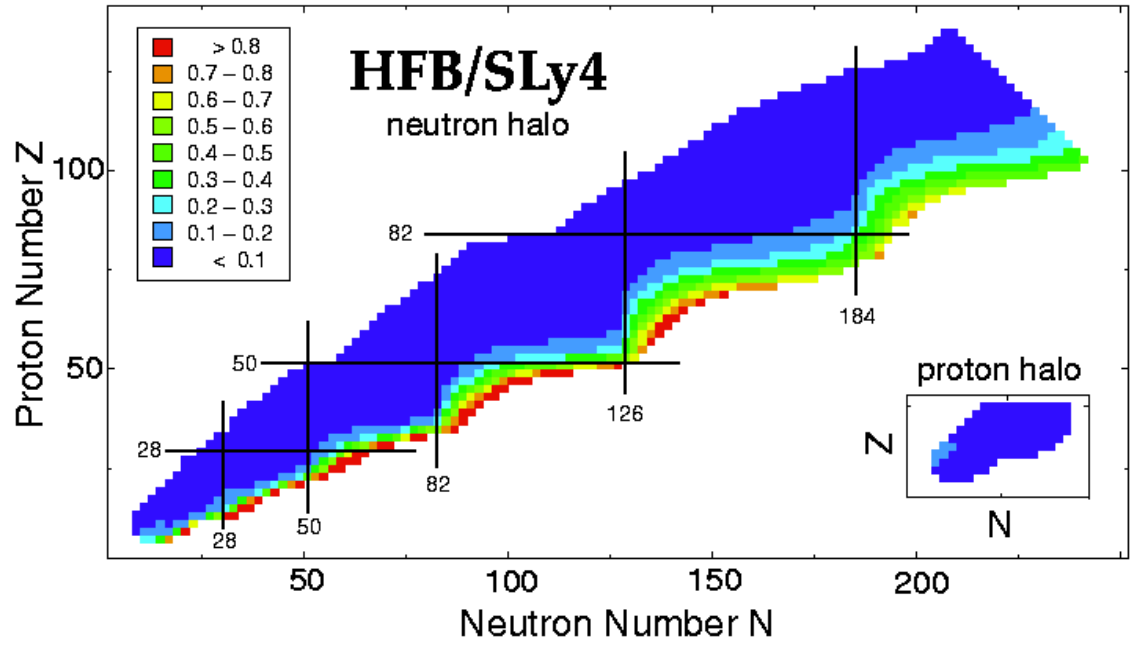}
\end{center}
\caption{Neutron halo parameters (\protect\ref{rhalo})
 calculated in the spherical HFB/SLy4 model for
two-particle stable even-even nuclei. The inset
shows proton halo parameters for very light nuclei.
}
\label{halon_SLy4}
}
\end{figure}
\myclearpagew

\begin{figure}
\vbox{
\begin{center}
\leavevmode
\epsfxsize=\myplotsizew
\epsfbox{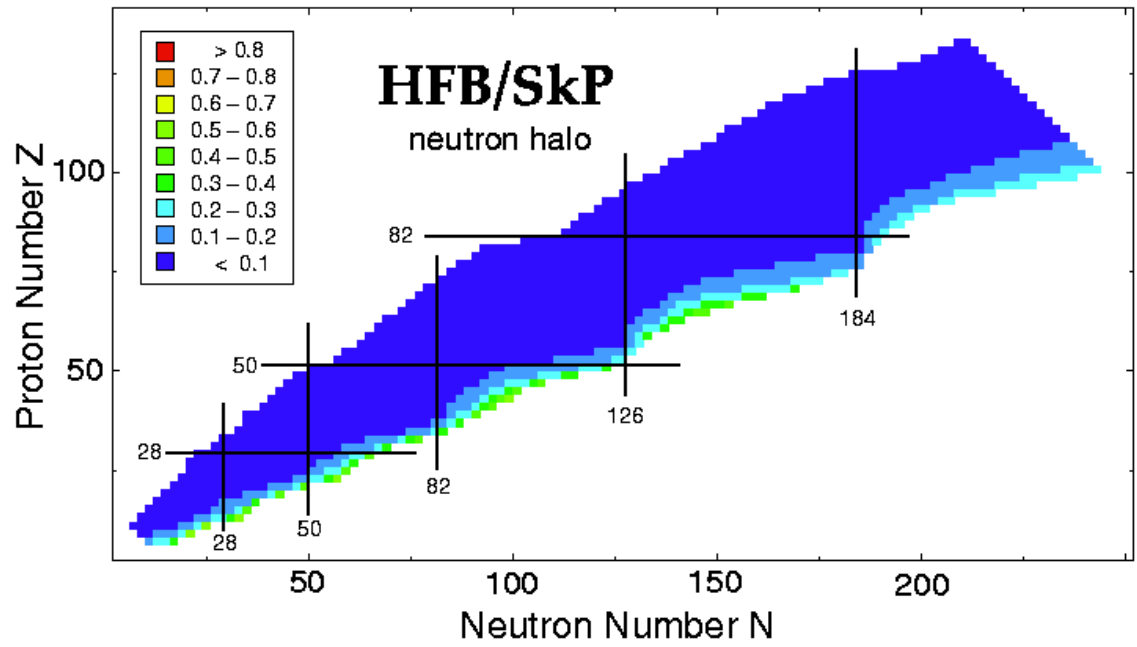}
\end{center}
\caption{Neutron halo parameters (\protect\ref{rhalo})
 calculated in the spherical HFB/SkP model for
two-particle stable even-even nuclei.
}
\label{halon_SkP}
}
\end{figure}
\myclearpagew

\begin{figure}
\vbox{
\begin{center}
\leavevmode
\epsfxsize=\myplotsizew
\epsfbox{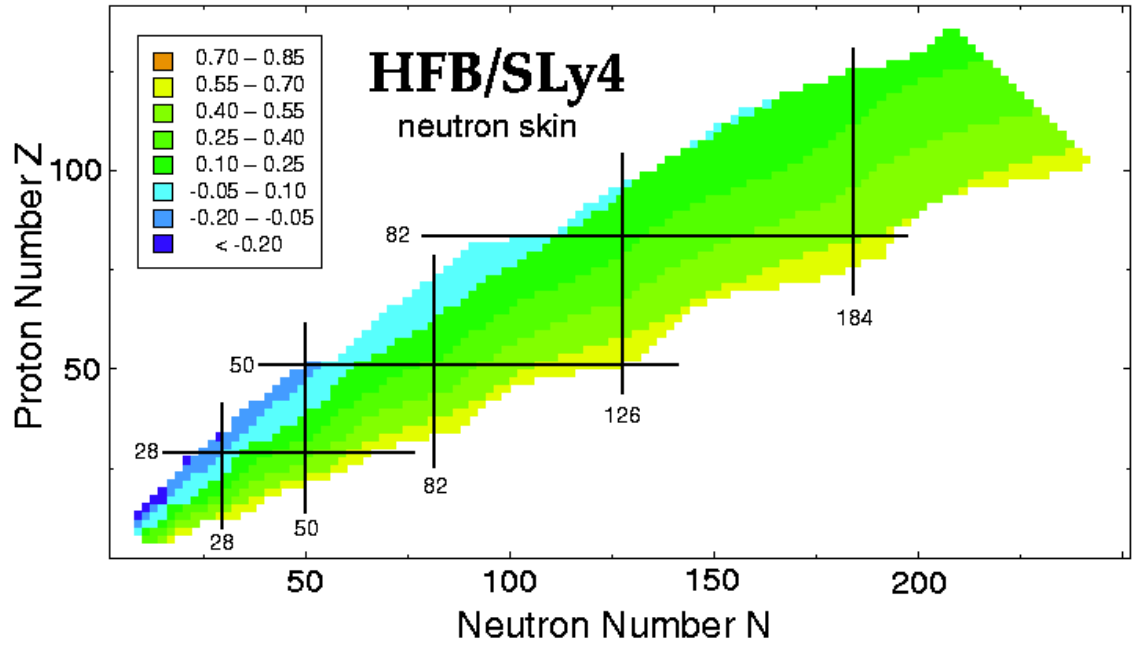}
\end{center}
\caption{Skin parameters (\protect\ref{rskinn})
 calculated in the spherical HFB/SkP model for
two-particle stable even-even nuclei.
}
\label{skins_SLy4}
}
\end{figure}
\myclearpagew

\begin{figure}
\vbox{
\begin{center}
\leavevmode
\epsfxsize=\myplotsizew
\epsfbox{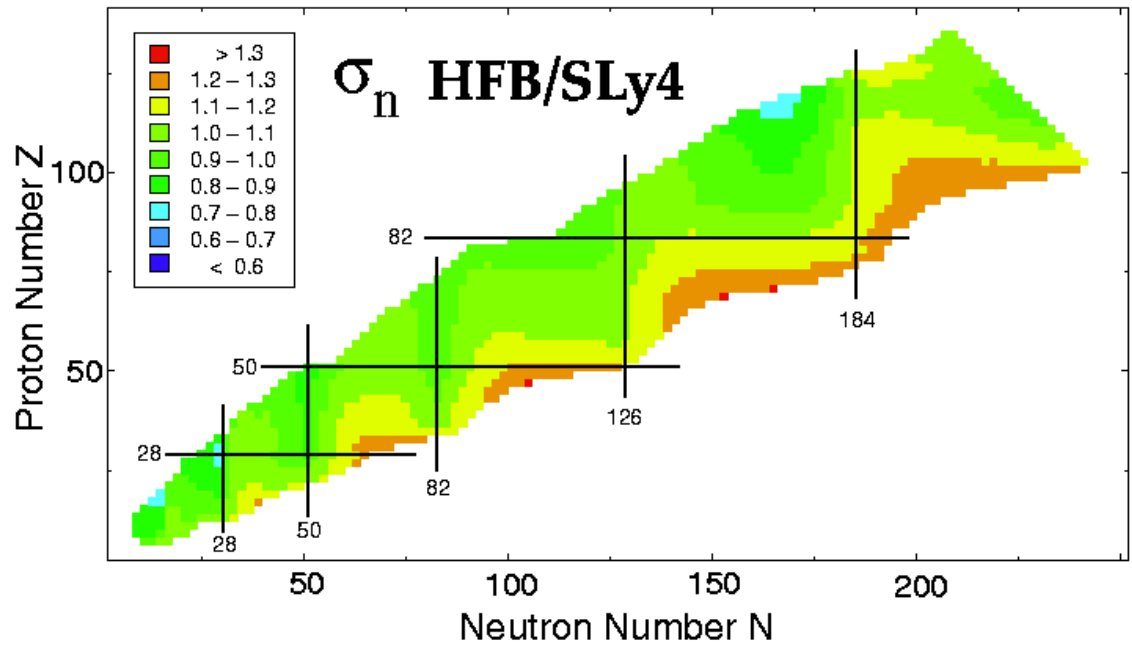}
\end{center}
\caption{Neutron surface thickness
 calculated in the spherical HFB/SkP model for
two-particle stable even-even nuclei.
}
\label{sigman}
}
\end{figure}
\myclearpagew

\begin{figure}
\vbox{
\begin{center}
\leavevmode
\epsfxsize=\myplotsizew
\epsfbox{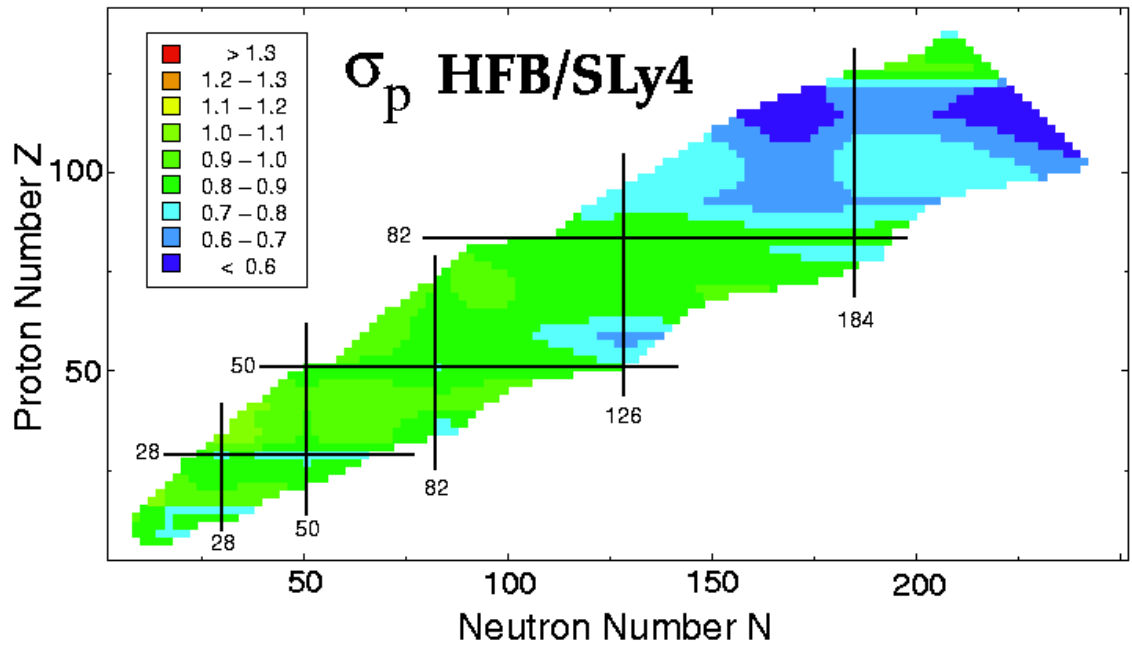}
\end{center}
\caption{Same as in Fig.~\protect\ref{sigman}, except for the
 proton surface thickness.
}
\label{sigmap}
}
\end{figure}
\myclearpagew

\begin{figure}
\vbox{
\begin{center}
\leavevmode
\epsfxsize=\myplotsizew
\epsfbox{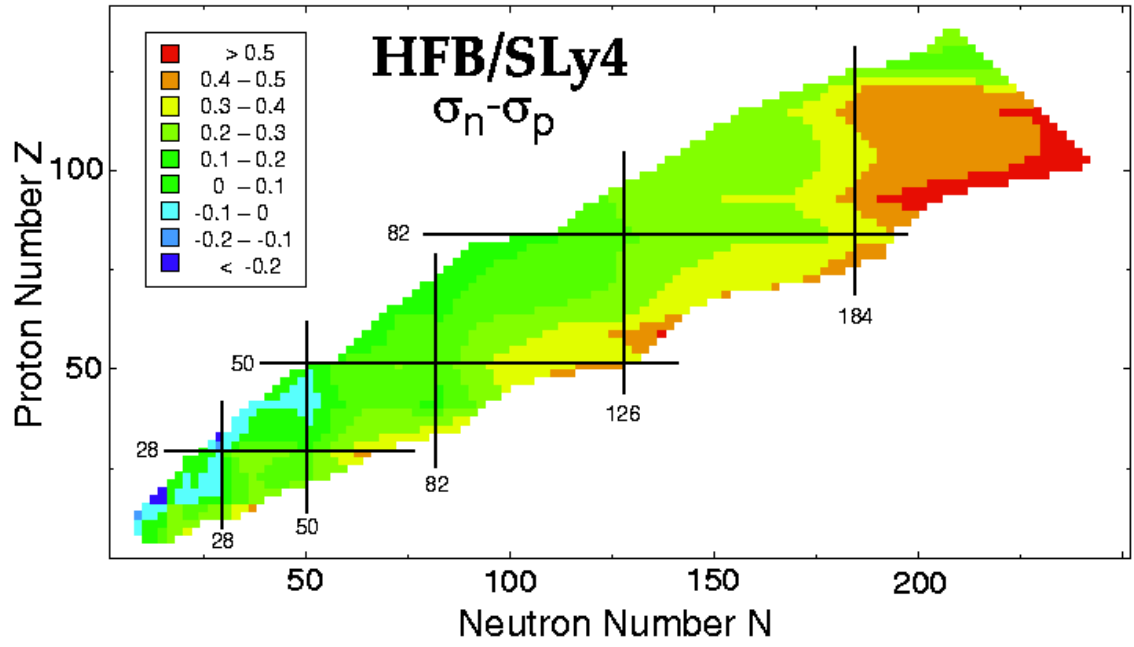}
\end{center}
\caption{Same as in Fig.~\protect\ref{sigman}, except for the
 difference $\sigma_n - \sigma_p$.
}
\label{deltasigmas}
}
\end{figure}
\myclearpagew

\narrowtext

\begin{figure}
\vbox{
\begin{center}
\leavevmode
\epsfxsize=\myplotsize
\epsfbox{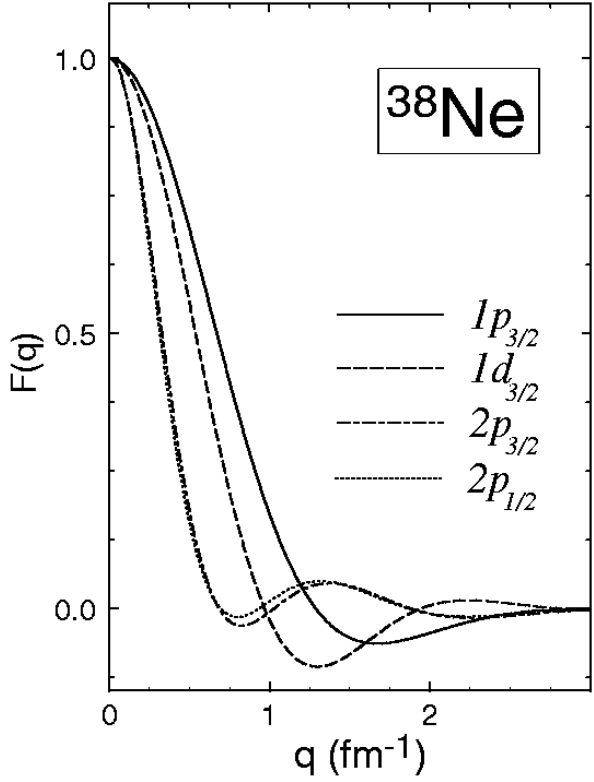}
\end{center}
\caption{Form factors of canonical RHB/NL3 single-neutron orbitals
1$p_{3/2}$, 1$d_{3/2}$, 2$p_{3/2}$, and 2$p_{1/2}$ in $^{38}$Ne.
}
\label{ffwfall}
}
\end{figure}
\myclearpage

\begin{figure}
\vbox{
\begin{center}
\leavevmode
\epsfxsize=\myplotsize
\epsfbox{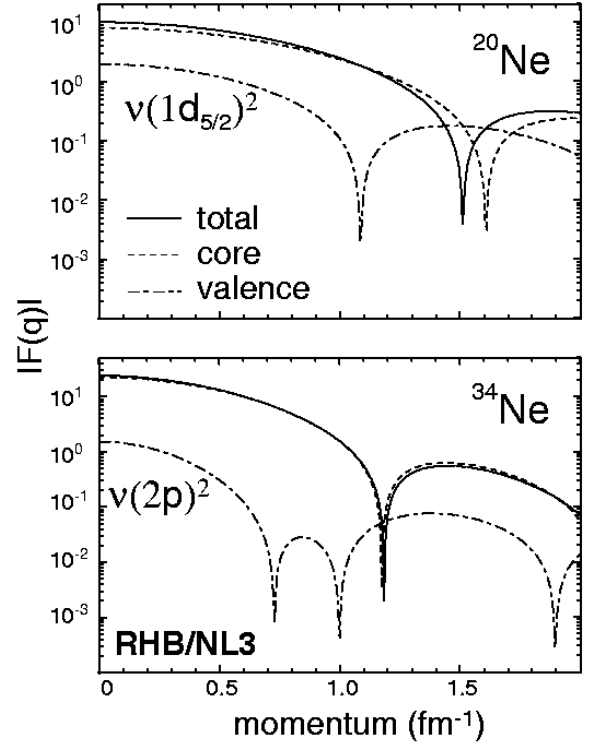}
\end{center}
\caption{Total (solid line), core (dashed line), and valence (dash-dotted line)
 neutron form factors  for $^{20}$Ne (top)
and  $^{34}$Ne (bottom). The valence space is here defined to be given by
two $1d_{5/2}$ neutrons in $^{20}$Ne and all the occupied $2p$ states
in $^{34}$Ne. The form factors at $q$=0 are normalized to the corresponding
neutron numbers
(e.g., $F_{\rm tot}(0)$=$N$).
}
\label{ff2034}
}
\end{figure}
\myclearpage

\begin{figure}
\vbox{
\begin{center}
\leavevmode
\epsfxsize=\myplotsize
\epsfbox{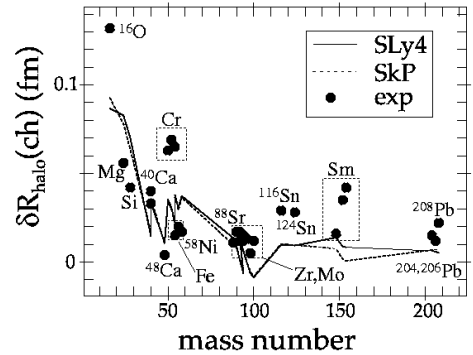}
\end{center}
\caption{
Comparison of experimental \protect\cite{[Fri99]} and theoretical charge halos
for a selection of stable nuclei. The spherical theoretical results are produced with
the two Skyrme parametrizations SLy4 and SkP used throughout this paper.
The estimated experimental errors are  about 0.03\,fm.
}
\label{halo_ch}
}
\end{figure}
\myclearpage

\end{document}